# Dynamic Bandwidth Allocation in Small-Cell Networks: An Economics Approach

Lin Cheng, Bernardo A. Huberman

Next-Gen Systems, CableLabs, Louisville, CO 80027, USA

*Abstract* — We propose and experimentally demonstrate a bandwidth allocation method based on the comparative advantage of spectral efficiency among users in a multi-tone small-cell radio access system with frequency-selective fading channels. The method allocates frequency resources by ranking the comparative advantage of the spectrum measured at the receivers ends. It improves the overall spectral efficiency of the access system with low implementation complexity and independently of power loading. In a two-user wireless transmission experiment, we observe up to 23.1% average capacity improvement by using the proposed method.

## I. INTRODUCTION

Multi-tone access systems, such as OFDMA, have dominated in wireless communications for decades thanks to their advantages in frequency-selective fading channels. With multiple subcarriers in the frequency domain, these systems also improve spectral efficiency by power loading and dynamic bandwidth/subcarrier allocation.

Optimal bandwidth allocation to multiple users with multiple subcarriers is a non-convex and NP-hard problem. The optimal solution is approximated by employing Lagrangian relaxation methods [1]. There are also other methods that can further reduce the complexity and iterations [2][3]. However, all these methods incorporate the constraints, including both power and bandwidth criteria, into the objective function. Thus the complete algorithm needs to be executed every time the constraints, e.g. users' bandwidth demands, change over time. This significantly increases implementation complexity and compromises the fluidity of the timeliness of bandwidth allocation.

In this paper, we adopt the concept of comparative advantage from macro-economics [4], and apply it among users in an OFDMA system. By noting the comparative advantage of spectral efficiency among users, the method reduces the opportunity cost during allocation and thus improves the overall efficiency and capacity of the access system. Most importantly, the method prioritizes the subcarriers regardless of the constraints. As a result, the algorithm needs to run much less frequently than other methods and therefore reduces the complexity and improves the dynamicity of allocation. Moreover, the optimization of subcarrier allocation is independent from power loading. Standalone power loading algorithms can be simply applied afterwards. In addition, considering the trend of cell densification in 5G and beyond, the method is tailored for small cells where a small number of concurrent high-bandwidth multi-user communications occur.

## II. COMPARATIVE ADVANTAGE

We first use a two-user two-subcarrier example to explain the concept of comparative advantage. The two subcarriers denoted as *A* and *B* have the same bandwidth and can be allocated to both user 1 and 2. On each subcarrier, under a certain power loading, the two users have different spectral efficiencies for carrying information because their channel conditions are different.

We want to know how to assign these two subcarriers to the two users so that we can have the highest overall efficiency. Assume both users want to use subcarrier *A* because they both perform better on *A* than *B*. In other words, subcarrier *A* has absolute advantage over subcarrier *B* for both users. However, we need to perform a relative comparison to decide if we should first assign user 1 or 2 to subcarrier *A* before we assign the leftover of subcarrier *A*, if there is any, after the user finishes occupying it, to the other user. The criterion for this comparison is called comparative advantage [4][5]. Let $\eta_{u,i}$ represent the spectral efficiency of subcarrier $i$ at serving user $u$. In this two-user two-subcarrier example, user 1 has comparative advantage at subcarrier *A* if

$$\frac{\eta_{1,A}}{\eta_{2,A}} > \frac{\eta_{1,B}}{\eta_{2,B}}. \qquad (1)$$

In other words, user 1 should always choose subcarrier *A* and user 2 should always choose subcarrier *B* as their first options when communications to both users are initiated. This is proven in [5].

To generalize the concept of comparative advantage to more than two subcarriers, we assume *N* subcarriers in a channel are shared by two users. The measured SNR (in linear) values of the two users under a certain power loading at the $i$-th subcarrier are denoted as $\gamma_{1,i}$ and $\gamma_{2,i}$, $1 \leq i \leq N$, respectively. The Shannon spectral efficiency of the two users at the $i$-th subcarrier are $\eta_{1,i} = \log(1 + \gamma_{1,i})$ and $\eta_{2,i} = \log(1 + \gamma_{1,i})$, $1 \leq i \leq N$, respectively. The method

ranks the ratios $\eta_{1,i}/\eta_{2,i}$, $1\leq i\leq N$, in a descending order and obtains the sequence of

$$\frac{\eta_{1,i_1}}{\eta_{2,i_1}} > \frac{\eta_{1,i_2}}{\eta_{2,i_2}} ... > \frac{\eta_{1,i_N}}{\eta_{2,i_N}} . \qquad (2)$$

The system selects the first $m$ subcarriers $i_1, i_2 ... i_m$ from the equation such that

$$\frac{\eta_{1,i_1}}{\eta_{2,i_1}} > \frac{\eta_{1,i_2}}{\eta_{2,i_2}} ... > \frac{\eta_{1,i_m}}{\eta_{2,i_m}} > T , \qquad (3)$$

where $T\geq 1$ is the threshold of comparative advantage that the system uses in its allocation. The system allocates these $m$ subcarriers to user 1 instead of user 2. Similarly, for the last $n$ subcarriers that

$$\frac{\eta_{2,i_N}}{\eta_{1,i_N}} > \frac{\eta_{2,i_{N-1}}}{\eta_{1,i_{N-1}}} ... > \frac{\eta_{2,i_{N-n+1}}}{\eta_{1,i_{N-n+1}}} > T , \qquad (4)$$

the system allocates them to user 2 instead of user 1 on account of their comparative advantage. For the subcarriers in the middle of equation (2), $i_{m+1}, i_{m+2} ... i_n$, the system has the freedom of allocating them to either user 1 or 2 subject to the allocation criteria without significantly compromising spectral efficiency.

To make the solution independent from power loading, let's further use the measured channel response, $h_{1,i}$ and $h_{2,i}$, as the indicators of comparative advantage. Equation (2) and (3) can be replaced by

$$\left|\frac{h_{1,i_1}}{h_{2,i_1}}\right| > \left|\frac{h_{1,i_2}}{h_{2,i_2}}\right| ... > \left|\frac{h_{1,i_m}}{h_{2,i_m}}\right| > T' \qquad (5)$$

and

$$\left|\frac{h_{2,i_N}}{h_{1,i_N}}\right| > \left|\frac{h_{2,i_{N-1}}}{h_{1,i_{N-1}}}\right| ... > \left|\frac{h_{2,i_{N-n+1}}}{h_{1,i_{N-n+1}}}\right| > T' , \qquad (6)$$

respectively. This is due to the fact that there always exists a $\delta > 0$ for a certain SNR range such that

$$\frac{\eta_{1,j}}{\eta_{2,j}} - \frac{\eta_{1,k}}{\eta_{2,k}} = \frac{\log(1+\gamma_{1,j})}{\log(1+\gamma_{2,j})} - \frac{\log(1+\gamma_{1,k})}{\log(1+\gamma_{2,k})} > 0 , \qquad (7)$$

$\forall 1 \leq j, k \leq N$, if

$$\frac{\gamma_{1,j}}{\gamma_{2,j}} - \frac{\gamma_{1,k}}{\gamma_{2,k}} = \frac{p_j/n_1}{p_j/n_2}\left|\frac{h_{1,j}}{h_{2,j}}\right| - \frac{p_k/n_1}{p_k/n_2}\left|\frac{h_{1,k}}{h_{2,k}}\right| > \delta , \qquad (8)$$

where $p_j$ and $p_k$ are the power loading coefficients on subcarrier $j$ and $k$, and $n_1$ and $n_2$ are the noise power at the two users. (8) is simplified to

$$\left|\frac{h_{1,j}}{h_{2,j}}\right| - \left|\frac{h_{1,k}}{h_{2,k}}\right| > \frac{n_1}{n_2}\delta > 0 . \qquad (9)$$

Note that in (9) power loading is cancelled and the ratios are only about the channel response.

In a more general case where there are more than two users, the users are first clustered into two groups, either randomly or based on users' frequency response. The average response of each group is calculated and used as the subject of advantage comparison. The results from the comparison between the two groups are ranked following (5) and (6). Based on the ranking, subcarriers are then allocated to the two groups accordingly. Within each group, we repeat the same procedure until there is only one user in each group. The average total computation complexity is $\mathcal{O}(UN \log N)$, where $U$ is the number of users. It is worth mentioning that the effectiveness of the method decreases as the number of users increase, making the method more suitable for small cells where only a few users concurrently consume a large bandwidth.

III. EXPERIMENTAL VERIFICATION

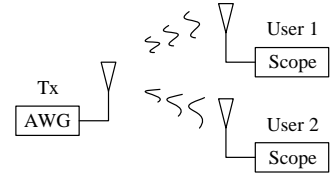

Fig. 1.  Wireless transmission setup.

We verified the feasibility and improvement of the method by performing a wireless transmission experiment in an office environment. The downlink transmitter (Tx) is an arbitrary waveform generator (AWG) and the two users are oscilloscopes, as shown in Fig. 1. The two users share the same 90-MHz downlink channel centered at 3.75 GHz (C-band licensed by special temporary authorization) with subcarrier spacing of 60 kHz. The channel response is measured at the user side and shown in Fig. 2. Because of multipath effects, the two users have different response and therefore different efficiencies over the frequency range. As movement inside the testing environment is limited, the response is considered consistent, i.e. slow fading channel, during the testing.

The method first estimates the spectral efficiency at the 125 resource blocks (each block contains 12 subcarriers) by using the channel response on each resource block, and then ranks the 125 ratios between the response of the two users as described by (5) and (6). The ratios are plotted in Fig. 3 indexed by the re-ranked resource blocks. Assuming $T' = 1.1$, the red curve in Fig. 3 tells us that the lower 63 resource blocks, i.e. the lower 756 subcarriers after re-ranking, should be allocated to user 2, while the blue curve, i.e. the inverse of the red curve, tells us that the higher 71 blocks after re-ranking should be allocated to user 1. The 7 blocks in between should not bring significant difference in terms of capacity no matter how they are allocated.

To see the improvement brought by the proposed method, in Fig. 4 we demonstrate the tradeoff of capacity between the two users that share the same 90-MHz channel.

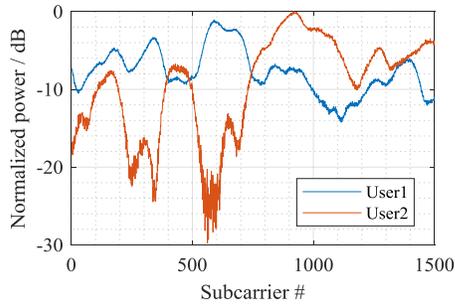

Fig. 2. Channel response measured at user 1 and 2 sharing the same 90-MHz downlink channel.

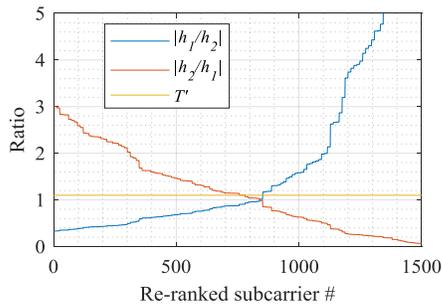

Fig. 3. Ratio of PSD after re-ranking resource blocks of subcarriers.

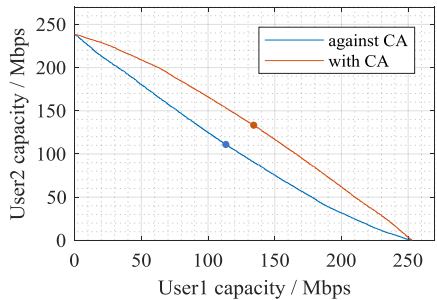

Fig. 4. Capacity of two users under allocation with and against comparative advantage (CA).

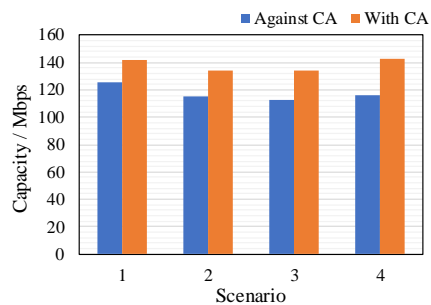

Fig. 5. Capacity improvement by using comparative advantage at four different pairs of user locations.

When the subcarriers are randomly assigned to the two users, the tradeoff curve varies in the space between the blue curve and the yellow curve. Over long-term averages, it is a straight line that connects the full-channel capacities of the two users – 250 and 240 Mbps in this example. However, there are situations where the system deteriorates to a convex curve, or even against comparative advantage, as shown by the blue curve. By using the proposed method, the system guarantees that the bandwidth allocation is always operating along the red curve, i.e. the best comparative advantage. Notice that when the two users have equal capacity, the proposed method provides 19% higher throughput than the worst case, by comparing the two dots in Fig. 4. Moreover, once the red tradeoff curve is derived, the system does not need to run the algorithm every time the allocation criteria change. The system only needs to renew the curve when there is obvious change on frequency response. Different response leads to different comparative advantages. We measure the improvement brought by comparative advantage by putting the two users in four pairs of locations. The results are shown in Fig. 5. The corresponding improvements are 13.4%, 16.7%, 19%, and 23.1%.

It is also worth noting that this method can be applied to both uplinks and downlinks although the example aforementioned is for downlinks only.

## IV. CONCLUSION

We proposed a bandwidth allocation method based on comparative advantage among users in a small-cell radio access system with frequency-selective fading channels. It allocates subcarriers by ranking the comparative advantage of the channel response measured at receiver(s) and improves the spectral efficiency of the access system. The algorithm requires a low implementation complexity and is independent of power loading. In experiments involving two users, we observed up to 23.1% improvement on capacity by using the proposed method.